\documentclass[11pt]{JHEP3}
\usepackage{multirow,colordvi}
\usepackage{latexsym}
\usepackage{epsf}
\usepackage{amssymb}
\usepackage{amsmath,amssymb,bbm}
\usepackage{multirow,colordvi}
\usepackage{epsf}
\usepackage{epsfig}
\usepackage{graphics}

\def\2{\frac12}
\def\4{\frac14}

\def\a{\alpha}

\def\D{\Delta}

\def\th{\theta}

\def\l{\lambda}

\def\m{\mu}
\def\n{\nu}

\def\r{\rho}

\def\S{\Sigma}
\def\t{\tau}
\def\f{\phi}
\def\F{\Phi}

\def\be{\begin{equation}}
\def\ee{\end{equation}}
\def\bea{\begin{eqnarray}}
\def\eea{\end{eqnarray}}
\author{Wissam A. Chemissany, Mees de Roo, \\Centre for
Theoretical Physics, University of
Groningen, Nijenborgh 4, 9747 AG Groningen, The Netherlands\\
\email{W.Chemissany, M.de.Roo@rug.nl}}
\author{Sudhakar Panda\\
Harish-Chandra Research Institute \\
Chatnag Road, Jhusi, Allahabad 211019, India\\
E-mail: \email{panda@mri.ernet.in}}

\title{Thermodynamics of Born-Infeld Black Holes}

\preprint{
UG-08-10\\
}


\abstract{We discuss the horizon structure for Born-Infeld 
black holes, in the context of Einstein-Born-Infeld gravity.
We show that the entropy function formalism agrees with a direct 
calculation of the entropy. With the entropy function 
formalism we 
also obtain the entropy when 
an axion-dilaton system 
as well as gravitational derivative corrections are included. 
}

\begin{document}

\section{Introduction\label{Intro}}

The work of Wald \cite{Wald:1993nt} has led to great progress in
the understanding of the entropy of black holes.
This approach is valid in
theories with invariance under general coordinate invariance,
and has been applied in particular to actions with
gravitational higher-derivative corrections (see \cite{Mohaupt:2007mb}
for an overview). For extremal black holes the entropy
formalism (\cite{Sen:2005wa}, see also \cite{Sen:2007qy}) 
can be applied. This method depends only on the action of
the theory being considered, in particular, it  
does not require the explicit black hole solution.

In this paper we emphasize higher-derivative terms in the 
matter contributions to the action. We are especially interested in
the coupling of gravity and Born-Infeld electromagnetism. This
higher-derivative version of Maxwell's theory leads to 
a generalization of the Reissner-Nordstrom black hole. The 
corresponding explicit black-hole solutions
have been known  for a long time, see, e.g,
\cite{Demianski:1986wx, Wiltshire:1988uq, Clement:2000ue, 
Gibbons:1995cv, Clement:2000ue, Breton:2002it, Breton:2007bza}. 
Born-Infeld theory is interesting
for many, not unrelated, reasons: it satisfies electric-magnetic
duality invariance, a property which also holds in the presence of
an axion-dilaton system \cite{Gibbons:1995ap}, Dirac-Born-Infeld theories, which
include additional scalars, correspond to the low-energy limit
of D-branes \cite{Andreev:1988cb,Leigh:1989jq} as well as
heterotic and Type I string theories. 
Also there is an interest for cosmological applications,
see \cite{Banados:2008rm} and references therein. Different aspects of the
Born-Infeld black holes have been also studied in, e.g., 
\cite{Rasheed:1997ns,Tamaki:2000ec,Tamaki:2001vv,Tamaki:2003pt,
Yazadjiev:2001uy,Chandrasekhar:2006ic,Chandrasekhar:2006zw,Stefanov:2007qw}.

We  focus on the black hole solutions of Einstein-Born-Infeld
(EBI) theory, and the extension that includes an axion and dilaton field
in an $SL(2,R)$ invariant way (EBIDA) \cite{Gibbons:1995ap}. In EBI
theory, where the exact black hole solution is available, we
can compare the result for the entropy by a direct calculation
using the solution, to the entropy function calculation 
for the extremal case. In EBIDA theory there is to the best of
our knowledge no exact solution available. In this case
we start from the black hole solution of 
Maxwell electromagnetism with axion and dilaton \cite{Kallosh:1993yg}.
In the case of two or more vector fields 
there is an extremal limit and the entropy formalism
can again be compared to the results from the solution.
In the EBIDA case the entropy
can only be obtained by the entropy function formalism. 
For completeness we extend these results to include also 
gravitational higher-derivative terms.

This paper is organized as follows. In Section \ref{EBI} we 
discuss the black hole solution in  EBI
electrodynamics and its horizon structure. We obtain thermodynamic 
quantities directly using the solution, and also from the 
near-horizon limit and the entropy function formalism. 
The results of the EBIDA case can be found in Section
\ref{EBIDA}, and their extension with higher derivative $R^2$ terms 
are in Section \ref{DC}. 
Our conclusions are in Section \ref{Conc}. In an Appendix
we include the equations of motion for the EBIDA case.

\section{Born-Infeld black holes\label{EBI}}

It has been known for a long time that the Reissner-Nordstrom
solution to the Einstein-Maxwell system can be extended to
the Einstein-Born-Infeld case 
\cite{Demianski:1986wx,Wiltshire:1988uq,Gibbons:1995cv,Breton:2007bza}.
The Einstein-Born-Infeld Lagrangian is of the form
\be
   {\cal L}_{EBI}= \sqrt{-\det{g}}R + \frac{4}{b^2}\left(
      \sqrt{-\det\ g} - \sqrt{-\det{(g+bF)}} \right)\,.
\label{actionEBI}
\ee
The spherically symmetric static solution with electric charge $q$ and
magnetic charge $p$ is of the form
\bea
  ds^2 &=& -G(r)dt^2 + \frac{dr^2}{G(r)} +
             r^2(d\th^2+\sin^2\th d\phi^2)\,,
\label{generalmetric}
\\
  F_{rt} &=&\frac{q}{\sqrt{r^4+a^4}}\,,\quad 
  F_{\th\f} = p\sin\th\,,
\label{solBI}
\\
  a^2 &=& b\sqrt{q^2+p^2}\,.
\label{defa}
\eea
with
\bea
   G(r) &=& 1-\frac{2m}{r} + \frac{q^2}{r^2}\left(
     \frac{4rg(r)}{3} +
     \frac{2r^4}{3a^4}\left(1 - \sqrt{1+\frac{a^4}{r^4}}\right)
     \right)\,,
\\
   g(r) &=& \frac{1}{2a}
      \left(-F(\varphi,\tfrac{1}{2}) + 2K(\tfrac{1}{2}\right)\,,\quad
   \,\,(0<\varphi\le \pi/2)
\\
    &=& \frac{1}{2a}\,F(\pi-\varphi,\tfrac{1}{2})\,,
   \phantom{+2K(\tfrac{1}{2}}\quad\quad (\pi/2\le\varphi<\pi)\,,
\\
   r &=& a\tan(\varphi/2)\,,
\label{solBIG}
\eea
where $F$ and $K$ are incomplete and complete elliptic integrals
of the first kind.

We will discuss some general properties of the solution 
(\ref{generalmetric}-\ref{solBIG}).
It is then convenient to set $p=0$. Magnetic charges can always be 
reinstated by using the duality property of Born-Infeld 
electrodynamics. 

For $b=0$ we recover the Reissner-Nordstrom solution. If in addition
$q=0$ the function $G(r)$ has one zero and we find the Schwarzschild
solution with horizon at $r_H=2m$. For $0<q<m$ $G(r)$ has two zeros at
$m\pm\sqrt{m^2-q^2}$. For $q=m$ these
two horizons coalesce at $r_H=m$, for $q>m$ there is no black hole.

\vspace*{-6.5cm}
\begin{figure}[h]
\includegraphics[scale=0.8]{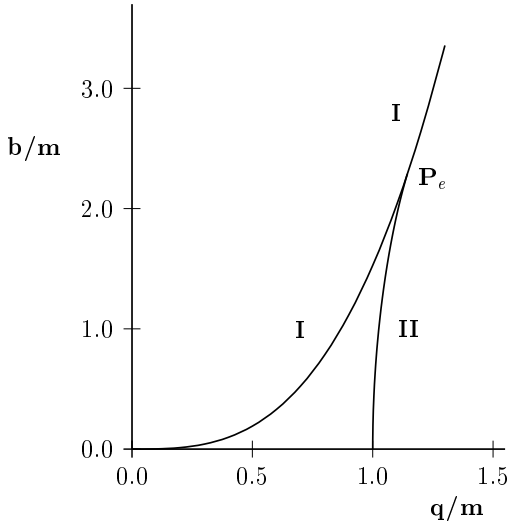}\vspace{-10.5cm}
\caption{The horizon structure of Born-Infeld black holes
as a function of the parameters $b$ and $q$. To the
left of curve I there is one horizon, in the area between curves
I and II there are two horizons, that coalesce on II. To the
right of II there are no horizons.}
\label{horizonfigure}
\end{figure}

For arbitrary $b>0$ the following properties are obtained
(see Figure 1).
The analogue of the Schwarzschild solution is found for
\be
   b > \frac{4q^3K^2(1/2)}{9m^2}\,.
\label{curveSG}
\ee
In the region where (\ref{curveSG}) is satisfied $G(r)\to -\infty$ for $r\to 0$, it has one zero,
and increases to the asymptotic value $G(r)=1$ at $r\to\infty$. On 
the boundary I determined by (\ref{curveSG}) $G'(0)=0$, and
$G(0)=1-2q/b$.
On the contrary, for
\be
  b < \frac{4q^3K^2(1/2)}{9m^2}
\label{curveGL}
\ee
$G(r)\to +\infty$ for $r\to 0$, and 
there is is region where $G(r)$ has two zeros. 
In the $(q,b)$ plane this region is bounded above by 
(\ref{curveGL}), and bounded to the right by the curve II that
runs from $(m,0)$ (extremal Reissner-Nordstrom) to the point
\be 
P_e=(q_e/m,2q_e/m),\qquad q_e/m=\frac{3}{\sqrt{2}K(1/2)}=1.144\ldots\,,
\label{pointPe}
\ee 
where the curves I and II intersect
On the curve II $G(r)$ and $G'(r)$ vanish at the point
\be
   r_H=\sqrt{q^2-\tfrac{1}{4}b^2}\,,
\label{horext}
\ee
which corresponds to
the horizon of the extremal Born-Infeld black hole. So there is a
finite range of extremal black holes. The horizon (\ref{horext})
shrinks from $r_H=q$ at $b=0$ to $r_H=0$ at $b=2q$ $(P_e)$.
To the right of both curves I and II there are no real zeros of $G(r)$
and there is no black hole.

For a metric of the form (\ref{generalmetric}) the requirement
that $G(r_H)=0$ and $G'(r_H)=0$ (which 
corresponds to an extremal solution with horizon $r_H$) gives a near-horizon metric
of the form (we now normalize to $m=1$):
\be
   ds^2=\frac{2}{G^{\prime\prime}(r_H)}\left(-r^{\prime\, 2} dt^{\prime\, 2}
    + \frac{1}{r^{\prime\, 2}} dr^{\prime\, 2}\right)
  + r_H^2\left(d\th^2+\sin^2\th d\f^2\right)\,,
\label{NH-metric}
\ee
where $r=r_H+\l r'$, $t'=\tfrac{1}{2}\l t G^{\prime\prime}(r_H)$,
in the limit $\l\to 0$. For the Born-Infeld case we find
\be
   G^{\prime\prime}(r_H) = \frac{2}{q^2+\tfrac{1}{4}b^2}\,. 
\ee
with $r_H$ given in (\ref{horext}). The entropy of the 
extremal Born-Infeld black hole is the area of the horizon,
\be
    {\cal S} = 16\pi^2(q^2-\tfrac{1}{4}b^2)\,,
\label{Entsol}
\ee
where we have set the gravitation coupling equal to $G_N=1/16\pi$.

Let us now apply the entropy function formalism \cite{Sen:2005wa}
to obtain the
entropy in this case. We start from
the near-horizon solution parametrized as
\bea
   ds^2_{NH} &=& v_1 (-r^2 dt^2+\frac{dr^2}{r^2})
              + v_2(d\th^2+\sin^2\th d\f^2)\,,
\nonumber\\
   F_{rt}&=& e \,,\qquad F_{\theta\phi}=p\sin\th\,.
\label{NHsol}
\eea
The entropy function is given by
\be
    {\cal E} = 2\pi\left( 16\pi q e - f(e,p,v_1,v_2)\right)\,,
\ee
where $f$ is the Lagrangian, evaluated in the near-horizon
limit (\ref{NHsol}), and integrated over the angles:
\be
     f(e,p,v_1,v_2) = \int d\th d\f {\cal L}_{NH}\,,
\ee
where in this case ${\cal L}_{NH}$ is (\ref{actionEBI}) 
evaluated for (\ref{NHsol}). One finds
\bea
  {\cal E} &=& 2\pi \bigg(
    16\pi q e - 4\pi \big(
   2(v_1-v_2) + \frac{4}{b^2}(v_1v_2 - 
    \sqrt{(v_2^2+b^2p^2)(v_1^2-b^2e^2)})\big)\bigg)\,.
\eea
The entropy is the value of ${\cal E}$ in the extremum with
respect to $e,v_1$ and $v_2$. We find
\be
    e=q\,,\qquad v_1=q^2+p^2+\tfrac{1}{4}b^2\,,\qquad 
                 v_2=q^2+p^2-\tfrac{1}{4}b^2\,,
\label{entropyEBIsol}
\ee
and 
\be
   {\cal E} = 16\pi^2\left(q^2+p^2-\tfrac{1}{4}b^2\right)\,, 
\ee
for the entropy, in agreement with the result obtained in
(\ref{Entsol}) from the explicit solution of the Born-Infeld black hole.

In the following sections we will extend this result to include
an axion-dilaton system as well as gravitational
higher-derivative corrections.

\section{Axion-dilaton\label{EBIDA}}

In this section we discuss the effect of adding an
axion-dilaton system to the results of section \ref{EBI}.
In this case there is, to our knowledge, no explicit solution
available for the Born-Infeld case.
We start with
an explicit solution for the Maxwell case, and will obtain
for that case the entropy both by the explicit calculation 
using the solution
and from the entropy formalism.

The Maxwell case 
was developed in a number of papers 
\cite{Shapere:1991ta,Garfinkle:1990qj,Ortin:1992ur,Kallosh:1993yg},
we will use the form given in the last reference. 
The action  for $N$ vector fields is given by
\bea
  {\cal L} &=& \sqrt{-\det g}\left( R + 2(\partial\F)^2
    + \tfrac{1}{2} {\rm e}^{4\F}(\partial a)^2 
    - {\rm e}^{-2\F} \sum_i \,F_iF_i\right) +
     a \sum_i \,F_i\, {}^*F_i \,.
\label{action-AD}
\eea
The solution is:
\bea
    ds^2 &=& -G(r)dt^2 + G(r)^{-1}dr^2
                  + R^2(r)(d\th^2+\sin^2\th d\f^2)\,,
\nonumber\\
    G(r) &=& (r-m-r_0)(r-m+r_0)/R^2(r)\,,\qquad
    R^2(r) = r^2 - \D^2 -\S^2\,,
\label{metricAD}
\\
    F_{i\,{rt}} &=& {\rm e}^{\F_0}\,
     \frac{q_i((r+\S)^2+\D^2)+2p_i\D r}{(r^2-\D^2-\S^2)^2}
    \,,\quad (i=1,\ldots,N)\,,
\\
    F_{i\,\theta\f} &=& -{\rm e}^{\F_0}\, p_i\,,\quad (i=1,\ldots,N)\,,
\label{EMAD}
\\
    {\rm e}^{2\F} &=& {\rm e}^{2\F_0}\,
        \frac{(r+\S)^2+\D^2}{r^2-\D^2-\S^2}
    \,,\qquad
    a(r) = a_0 - {\rm e}^{2\F_0}\,\frac{2\D r}{(r+\S)^2+\D^2}\,.
\label{ADAD}
\eea
The mass $m$, dilaton and axion charges $\S, \D$,
and the electric and magnetic charges $q_i$ and $p_i$
are subject to the following conditions:
\bea
   && r_0{}^2=m^2+\D^2+\S^2 - (P^2+Q^2)\ge 0 \,,
\nonumber\\
   && P\cdot Q + m\D = 0\,,
\nonumber\\
   && - P^2+Q^2  + 2m\S = 0\,,
\label{conditions-AD}
\eea
where
\be
   P^2\equiv\sum_i \,p_i^2,\ Q^2\equiv\sum_i\, q_i^2,\ 
   P\cdot Q\equiv\sum_i\, p_iq_i\,.
\ee
Before turning to the extremal case with $N$ vector fields 
we briefly consider this system of equations for $N=1$, i.e.,
one vector. Then the constraints imply:
\be
    \D^2+\S^2= \left(\frac{p_1^2+q_1^2}{2m}\right)^2\,,
\qquad r_0^2=\left(m-\frac{p_1^2+q_1^2}{2m}\right)^2\,.
\ee
We then find:
\be
   G(r)=\frac{r-2m +\frac{p_1^2+q_1^2}{2m}}{r+\frac{p_1^2+q_1^2}{2m}}\,,
\qquad
   R^2(r)= r^2 - \left(\frac{p_1^2+q_1^2}{2m}\right)^2\,.
\label{vec1sol}
\ee
This solution with a single horizon, when the axion field is absent,
is the one found in \cite{Garfinkle:1990qj}.
The curvature singularity is at $r_S=(p_1^2+q_1^2)/2m$. The solution
has a horizon at $r_H=2m-r_S$, if $r_H>r_S$, which corresponds to
$p_1^2+q_1^2<2m^2$. If $p_1^2+q_1^2=2m^2$ the horizon and curvature
singularity coincide. This charged axion-dilaton black hole 
with a single vector has no extremal limit.

Now we go back to (\ref{conditions-AD}) and the $N$-vector case.
Here the constraints imply:
\bea
   r_0{}^2 = \frac{1}{4m^2}\left( (2m^2-P^2-Q^2)^2 
          - 4 S^2 \right)\,,
  \qquad
\label{r0value}
\eea
where 
\be
   S = \sqrt{P^2Q^2 - (P\cdot Q)^2}\,.
\label{Svalue}
\ee
The curvature singularity is at $r_S{}^2=\D^2+\S^2$. The solution
becomes extremal for $r_0=0$, which leads to
\be
   2m^2=P^2+Q^2 \pm 2 S\,.
\label{mextremal}
\ee
The condition that the horizon for the extremal case is outside
the curvature singularity leads to
\bea
  (r_H)^2 &=& m^2 = \tfrac{1}{2} (P^2+Q^2 + 2 S )\,,
\label{rH1}
\eea
The near-horizon limit of the metric is
\be
  ds^2 = (m^2-\D^2-\S^2)\left(-\r^2 d\t^2+ \frac{d\r^2}{\r^2}\right)
        +(m^2-\D^2-\S^2)(d\theta^2+\sin^2\theta d\phi^2)\,,
\ee
with, for the case where
$m^2$ is given by (\ref{rH1}),
\be
  m^2 -\D^2-\S^2 = 2 S \,.
\ee
The metric exhibits the usual $AdS_2\times S_2$ symmetry.
The difference with (\ref{NH-metric}) is due to the fact that
the curvature singularity is not at $r=0$ in this case. The entropy
is given by
\be
    {\cal S} = 16\pi^2((r_H)^2 - (r_S)^2) = 32\pi^2 S \,.
\label{entropy-sol}
\ee

Now we will treat the same case by the entropy formalism.
The entropy function  is
\bea
   {\cal E} &=& 2\pi\bigg(
    16\pi Q\cdot E - 4\pi
    \bigg( 
      2(v_1-v_2) -4a P\cdot E- 
     2{\rm e}^{-2\F}\,\frac{P^2 v_1^2- E^2 v_2^2}{v_1v_2}\bigg)\bigg)\,.
\label{entFAD}
\eea
Here $E=(e_1,\ldots e_N)$ is the vector of electric fields 
$F_{i\,rt}$.
We require that ${\cal E}$ is extremal under variations of
$e_i,v_1,v_2,a$ and ${\rm e}^{-2\F}$. This gives the following
equations:
\bea
   && q_i+ap_i-{\rm e}^{-2\F}e_i v_2/v_1=0\,,
\label{e1eq}
\\
   && P\cdot E = 0\,,
\label{aeq}
\\
   && 2{\rm e}^{-2\F}\left(P^2 v_1/v_2 - E^2 v_2/v_1\right)=0\,,
\label{DilEq}
\\
   && 1-{\rm e}^{-2\F}\left(P^2 /v_2 + E^2v_2/v_1^2)\right)=0\,,
\label{v1Eq}
\\
   && 1-{\rm e}^{-2\F}\left(P^2 v_1/v_2^2 + E^2/v_1\right)=0\,,
\label{v2Eq}
\eea
The solution is:
\bea
   v_1&=&v_2= 2S\,,
\label{solv}
\\
   a&=& - {P\cdot Q}/{P^2}\,,
\label{sola}
\\
  {\rm e}^{-2\F} &=&  {S}/{P^2}\,,
\label{soldil}
\\
  e_i&=& {(P^2q_i-P\cdot Q p_i)}/{S}\,,
\label{sole}
\eea
If we substitute this back into the entropy function we find
\be
   {\cal E} = 32\pi^2 S\,,
\label{entropy-EF}
\ee
in agreement with (\ref{entropy-sol}).

To obtain information on the entropy for the Born-Infeld case we have to
use the entropy function formalism, since there is no explicit
solution available. 
The action reads
\bea
  {\cal L} &=& \sqrt{-\det g}\left( R + 2(\partial\F)^2
    + \tfrac{1}{2} {\rm e}^{4\F}(\partial a)^2\right) 
    + a \sum_i\, F_i\, {}^*F_i + \sum_i\, {\cal B}_i\,,
\label{action-ADBI}
\eea
with
\be
   {\cal B}_i = \frac{4}{b^2}\left(\sqrt{-\det g} 
      -\sqrt{-\det(g+b{\rm e}^{-\F}F_i)}
       \right) \,.
\label{BI-cont}
\ee
This satisfies the requirements of electric-magnetic duality
\cite{Gibbons:1995ap}.
In this case the entropy function is
\bea
  &&{\cal E_{BI}} = 2\pi\big(
    16\pi Q\cdot E - 4\pi 
    \big( 
      2(v_1-v_2) -4a P\cdot E + \sum_i\,{\cal B}_{NH\,i}
 \big)\big)\,,
\label{entFADBI}  
\eea
where the Born-Infeld contribution is now expressed in terms
of fields in the near-horizon limit: 
\be
   {\cal B}_{NH\,i} = \frac{4}{b^2}\left(v_1v_2 
  -\sqrt{(v_1^2-b^2e_i^2{\rm e}^{-2\F})
         (v_2^2+b^2p_i^2{\rm e}^{-2\F})}\right)
\label{BI-nhlim}
\ee
We obtain the following equations from the variation of 
${\cal E_{BI}}$:
\bea
 0&=&q_i+ap_i-\frac{{\partial\cal B}_{NH\,i}}{4\partial e_i}\,,
\label{eiBIeq}
\\
 0&=& P\cdot E\,,
\label{aBIeq}
\\
0 &=& \sum_i\,\frac{\partial{\cal B}_{NH\,i}}{\partial {\rm e}^{-2\F}}
\label{DilBIeq}
\\
 0&=& 2 + \sum_i\,\frac{\partial{\cal B}_{NH\,i}}{\partial v_1}
\label{v1BIeq}
\\
 0&=& -2 + \sum_i\,\frac{\partial{\cal B}_{NH\,i}}{\partial v_2}
\label{v2BIeq}
\eea
In the general case we have not found an explicit solution of these
equations. In an expansion in $b^2$ it is possible to find
a solution for the variables $e_i,\ v_1,\ v_2,\ a$ and ${\rm e}^{-2\F}$
to order $b^2$.
The corresponding entropy is
\be
    {\cal E}_{BI} = 16\pi^2 \left( 2S - 
  \frac{b^2}{16 (P^2)^2}\sum_i(e_{i}^2 + p_i^2)^2 + 
   {\cal O}(b^4) \right)\,,
\ee
where $e_i$ takes the value (\ref{sole}).

For the special case $N=2$, where $S=|p_1q_2-p_2q_1|$ 
we find
\bea
   v_1&=& 2|p_2q_1-p_1q_2|+\frac{b^2}{8}\,, 
   \quad v_2 = 2|p_2q_1-p_1q_2|-\frac{b^2}{8}\,,
\label{solBIv}
\\
   a&=& - \frac{p_1q_1+p_2q_2}{p_1^2+p_2^2}\,,
\label{solBIa}
\\
  {\rm e}^{-2\F} &=&  \frac{|p_2q_1-p_1q_2|}{p_1^2+p_2^2}\,,
\label{solBIdil}
\\
  e_1&=& \frac{p_2(p_2q_1-p_1q_2)}{|p_2q_1-p_1q_2|}\,,
\label{solBIe1}
\\
  e_2&=& -\frac{p_1(p_2q_1-p_1q_2)}{|p_2q_1-p_1q_2|}\,.
\label{solBIe2}
\eea
Except for the $b^2$-contribution in $v_1$ and $v_2$
this is the same solution as for the Maxwell case, see
(\ref{solv}-\ref{sole}). See also the Born-Infeld solution
without axion-dilaton, where there is a similar structure 
(\ref{entropyEBIsol}).
The entropy is:
\be
   {\cal E}_{BI} = 16\pi^2\left(2|p_2q_1-p_1q_2|-\frac{b^2}{8}\right)
\label{EntBI}
\ee
This simple solution, and the similarity with the results of Section 
\ref{EBI}, might be helpful in finding an explicit solution for the Born-Infeld 
case with axion and dilaton for $N=2$.

\section{Gravitational higher-derivative corrections\label{DC}}

In this section we will consider gravitational higher-derivative terms
of the form $R^2$. We will add to the Lagrangians
(\ref{actionEBI},\ \ref{action-ADBI}) terms of the form
\be
   {\cal L}_\alpha= \alpha{\rm e}^{-\F}\left(
     x R_{\m\n\l\r}R^{\m\n\l\r} 
    -4y R_{\m\n}R^{\m\n}
    + z R^2 \right)\,.
\ee
Here again it is difficult, if not impossible, to obtain exact solutions.
What we can do, is to work to order $\a$ \footnote{Of course, in
string theory $b$ is proportional to $\a$, so
one should also truncate the Born-Infeld system. However, there is no problem 
in keeping the complete $b$-dependence, which we therefore do.}.
In this case the result depends only on the solution at $\a^0$,
corrections to this solution
due to the higher-derivative terms contribute only to terms of 
order $\a^2$ and higher.

In the calculation of the entropy function we use:
\be
   R_{\m\n\l\r}R^{\m\n\l\r} = \frac{4(v_1^2+v_2^2)}{v_1^2v_2^2}\,,\quad
   R_{\m\n}R^{\m\n} = \frac{2(v_1^2+v_2^2)}{v_1^2v_2^2}\,,\quad
   R^2 = \frac{4(v_1-v_2)^2}{v_1^2v_2^2}\,.
\label{gravHD}
\ee
Working to order $\a$, and including axion and dilaton, we
find for the case $N=2$, 
\bea
 {\cal E} &=& 16\pi^2\left(2|p_2q_1-p_1q_2|-\frac{b^2}{8}\right) - {}
 \nonumber\\
&&
 \qquad  \frac{64\pi^2\alpha{\rm e}^{-\F_0}\left(
   (256 |p_2 q_1 - p_1 q_2|^2 (x - 2 y) + b^4 (x - 2 y + 2 z)\right)}
  {256 |p_2 q_1 - p_1 q_2|^2-b^4 }\,.
\eea
Here ${\rm e}^{-\F_0}$ corresponds to the solution (\ref{solBIdil}).
The Gauss-Bonnet combination $(x=y=z=1)$ depends on the charges
only through ${\rm e}^{-\F_0}$, and is independent of the
Born-Infeld parameter $b$.

\section{Conclusions\label{Conc}}

The main conclusion is that the entropy function formalism
works well in all cases considered in this paper:
Einstein-Born-Infeld black holes, and the various extensions
thereof, including in particular the axion-dilaton case.
In this last case we cannot compare with the result of an
exact solution, but the result of the entropy function
agrees, in various limits, with known results.

The simplicity of the EBIDA entropy for $N=2$ seems to suggest that
an explicit black hole solution might be obtained, however, 
we have not succeeded in this respect. An extension
of our work would be to include not only a general $R^2$
gravititonal higher-derivative term, but a complete order $\alpha'$
correction indicated by string theory. In specific cases,
such as the heterotic string, this can be pushed to higher orders in
$\alpha'$. It would be interesting to see how the entropy function 
formalism copes with such an extension.

\acknowledgments

We are grateful to Tomas Ort\'{\i}n, Sjoerd de Haan and
Ashoke Sen for useful discussions.
SP thanks  the Centre for Theoretical Physics in Groningen for their
hospitality. WC and MdR  are supported by the European
Commission FP6 program MRTN-CT-2004-005104 in which WC and MdR
are associated to Utrecht University.

\appendix

\section{Equations of motion\label{EOM}}

In this Appendix we present the equations of motion for
the Maxwell and Born-Infeld cases, corresponding to the 
action given in  section \ref{EBIDA}. The metric is
always of the form
\bea
    ds^2 &=& -G(r)dt^2 + G(r)^{-1}dr^2
                  + R^2(r)(d\th^2+\sin^2\th d\f^2)\,,
\label{metric-all}
\eea
We give the equations for a single vector field. 
The electric and magnetic fields appear as
\be
    F{}_{\,rt} = E(r)\,,\quad F{}_{\theta\phi} = M(r)\sin\theta\,.
\ee
The  equation for the vector field is of the form:
\bea
    0 &=& \left( M(r) a(r) + E(r){\rm e}^{-2\F(r)}X(r)\right)'\,,
\label{eom-vector}
\eea
where the prime indicates differentiation with repect to $r$,
and
\bea
     X(r) &=& R^2(r)+\frac{b^2}{2}{\rm e}^{-2\F(r)}
    \left( \frac{M^2(r)}{R^2(r)}+ E^2(r)R^2(r)\right)\,,
     \ {\rm Maxwell\ case,\ }b\to 0\,,
\nonumber\\
     X(r) &=& \sqrt{\frac{R^4(r){\rm e}^{2\F(r)}+b^2M^2(r)}
                         {{\rm e}^{2\F(r)}- b^2 E^2(r)}}\,,
    \qquad\qquad\qquad\qquad\ 
    {\text{\rm Born-Infeld\ case,}}\ b\ne 0\,.
\eea
The dilaton and axion equations of motion are respectively:
\bea
  0&=& -\left(2 G(r)R^2(r) \F(r)' \right)' +
         {\rm e}^{4\F(r)} G(r) (a(r)')^2 
       - 2{\rm e}^{-2\F(r)}\left( \frac{M^2(r)}{X(r)} 
                               -      E^2(r)X(r) \right)\,,
\label{eom-dilaton}
\\
  0&=& \left({\rm e}^{4\F(r)}G(r)R^2(r) a(r)'\right)'
      + 4  E(r)M(r) \,.
\eea
There are three independent components of the Einstein equations.
Taking convenient linear combinations of these gives the following
three equations of motion:
\bea
  0&=& {\rm e}^{4\F(r)} (a(r)')^2+ 4 (\F(r)')^2
        - \frac{1}{R^4(r)}
     \left( (R^2(r)')^2 - 2 R^2(r) (R^2(r))^{\prime\prime}
          \right)\,,
\label{eom-einstein1}
\\
  0&=&  -2 + (G(r)R^2(r)')' -\frac{4R^2(r)}{b^2}
     \left( 1 - \frac{X(r)}{R^2(r)}\right)\,,
\label{eom-einstein2}
\\
  0&=& -2 +  (G(r)R^2(r))^{\prime\prime}
        - \frac{4R^2(r)}{b^2}\left(
               2 - \frac{X(r)}{R^2(r)} -
                   \frac{R^2(r)}{X(r)} \right)\,.
\label{eom-einstein3}
\eea

In (\ref{eom-einstein3}) we can write
\be
   (G(r)R^2(r))^{\prime\prime} = (G(r)'R^2(r))'+ (G(r)R^2(r)')'\,,
\ee
which we can use to combine with (\ref{eom-einstein2}), giving
instead of (\ref{eom-einstein3}):
\be
   0 = (G(r)'R^2(r))' - \frac{4R^2(r)}{b^2}\left(
              1 - \frac{R^2(r)}{X(r)} \right)\,.
\label{eom-einstein3a}
\ee

\subsection{The Einstein-Born-Infeld black hole\label{BI}}

Here we obtain the solution discussed in section \ref{EBI}.
Thus we have Born-Infeld electromagnetism,
Einstein gravity, but no dilaton and axion. 
Solve equation (\ref{eom-einstein1}) by:
\be
    R^2(r) = r^2\,.   
\ee
Then we can solve (\ref{eom-vector}), which gives:
\be
    E(r)R^2(r) = Q\sqrt{1-b^2E(r)^2}\quad \to \quad
    E(r) = \frac{Q}{\sqrt{r^4+b^2Q^2}} = \frac{Q}{X(r)}\,.
\label{Esol}
\ee
There are two remaining equations (\ref{eom-einstein2})
and (\ref{eom-einstein3}):
\bea
  && -2 + 2 (rG(r))' = \frac{4r^2}{b^2}\left(
           1- \frac{Q}{r^2E(r)}\right)\,,
\label{BI-1}
\\
  && -2 + 2(rG(r))' + r(rG(r){}^{\prime\prime}
     = \frac{4r^2}{b^2}\left(
           2- \frac{Q}{r^2E(r)}-\frac{r^2E(r)}{Q}\right)\,.
\label{BI-2}
\eea
Note that (\ref{BI-1}) and (\ref{BI-2}) combine to
\be
     (rG(r)){}^{\prime\prime} = \frac{4r}{b^2}\left(
           1-\frac{r^2E(r)}{Q}\right)\,.
\label{BI-2a}
\ee
The solution for the vector field (\ref{Esol}) can be written as:
\be
    E(r) = - Qg(r)'\,,\quad 
    g(r)' \equiv -\frac{1}{\sqrt{r^4+b^2Q^2}}\,.
\ee
From (\ref{BI-2}) we then obtain
\be
   (rG(r)){}^{\prime\prime} = \frac{4r}{b^2} + 
                 \frac{4r^3 g(r)'}{b^2}\,,
\label{BI-3}
\ee
which we also obtain by differentiating 
(\ref{BI-1}). So  (\ref{BI-1}) is the integrated
form of (\ref{BI-2a}). We can then integrate 
(\ref{BI-1}) to
\be
   rG(r) = -2m + r + \frac{2r^3}{3b^2}
         \left(1-\frac{\sqrt{r^4+b^2Q^2}}{r^2}\right)
               + \frac{4Q^2g(r)}{3}\,,
\label{BI-4}
\ee
which is the solution (\ref{solBIG}).

\end{document}